\documentclass[aps,onecolumn,12pt,superscriptaddress,showpacs]{revtex4-1}

\usepackage{graphicx,color}
\usepackage{amsmath}
\usepackage{amssymb}

\renewcommand{\Re}{\mathop{\text{Re}}\nolimits}

\newcommand{\Tr}{\mathop{\text{Tr}}\nolimits}

\begin{document}

\title{Statistical Interparticle Potential \\of an Ideal Gas of Non-Abelian Anyons}

\author{Francesco Mancarella}
\affiliation{SISSA and INFN, Sezione di Trieste, via Bonomea 265, 
I-34136 Trieste, Italy}

\author{Andrea Trombettoni}
\affiliation{CNR-IOM DEMOCRITOS Simulation Center, Via Bonomea 265, I-34136 Trieste, Italy}
\affiliation{SISSA and INFN, Sezione di Trieste, via Bonomea 265, 
I-34136 Trieste, Italy}

\author{Giuseppe Mussardo}
\affiliation{SISSA and INFN, Sezione di Trieste, via Bonomea 265, 
I-34136 Trieste, Italy}
\affiliation{International Centre for Theoretical Physics (ICTP), 
Strada Costiera 11, I-34151, Trieste, Italy}

\begin{abstract}
We determine and study the statistical interparticle potential 
of an ideal system of 
non-Abelian Chern-Simons (NACS) particles, comparing our results with the 
corresponding results of an ideal gas of Abelian anyons. In the 
Abelian case, the statistical potential depends on the statistical parameter 
$\alpha$ and it has a ``quasi-bosonic'' behaviour 
for $0<\alpha<1/2$ (non-monotonic with a minimum) and a 
``quasi-fermionic'' behaviour for $1/2<\alpha<1$ 
(monotonically decreasing without 
a minimum). In the non-Abelian case the behavior of the 
statistical potential depends on the Chern-Simons coupling and 
the isospin quantum number: as a function of these two parameters, 
a phase diagram with quasi-bosonic, quasi-fermionic and bosonic-like regions is obtained and 
investigated. Finally, using the obtained expression for the statistical 
potential, we compute the second virial coefficient of the NACS gas, which correctly 
reproduces the results available in literature.

\begin{small}\textbf{Keywords:}\end{small} 
\begin{footnotesize} 
Fractional statistics;\, Anyon thermodynamics;\, Effective statistical potential; Chern-Simons theory \end{footnotesize}\\
\begin{small}\textbf{PACS number:}\end{small} 05.30.Pr\\
\begin{small}\textbf{Contact information:}\end{small}
\begin{small}Francesco Mancarella (corresp. author):\end{small} \begin{footnotesize} mancarel@sissa.it 
\end{footnotesize}

\end{abstract}

\maketitle

\section{INTRODUCTION}
A key property of the statistics of quantum systems in two space 
dimensions is provided by the possibility to display intermediate fractional 
statistics interpolating between bosons and fermions: the properties 
of anyons, the two-dimensional identical particles obeying 
fractional braiding statistics and carrying fractional charge, have been 
the subject of an intense and continuing interest 
\cite{Leinaas77,Wilczek90,Lerda92,Khare05,Nayak08}. 
Both Abelian and non-Abelian anyons (associated to respectively 
one-dimensional and irreducible higher-dimensional representations 
of the braid group) have been extensively studied both  
for their intrinsic interest and their connection with quantum Hall systems 
\cite{Nayak08,Yoshioka02,Jain07}. In particular there is an increasing interest in 
the investigation of the properties of non-Abelian anyons for their application 
to topologically fault-tolerant quantum information processing \cite{Nayak08}.

The ideal gas of anyons is also interesting for the phenomenon   
of statistical transmutation 
\cite{Wilczek90,Lerda92,Khare05}, i.e. the fact that one can treat 
non-interacting anyons as interacting bosons or fermions. The idea behind 
statistical transmutation is that one can alternatively consider the system 
as made of interacting particles with canonical statistics or 
of non-interacting particles, but obeying non-canonical statistics. This 
makes in general difficult the study of the equilibrium 
thermodynamical properties of the ideal anyon gases \cite{Khare05} and 
therefore any result that gives even qualitative informations on such ideal 
gases is valuable.

Dating back to the seminal work by Uhlenbeck and Gropper 
\cite{Uhlenbeck32}, a standard way to characterize the effects of the 
quantum statistics on the properties of ideal gases 
is provided by the determination of the so-called 
statistical potential. As detailed in textbooks \cite{Pathria72,Huang87}, 
one can show that the partition function (PF) of a gas of 
particles approaches - for sufficiently high temperatures - the PF 
of the classical gas (with the correct Boltzmann counting). 
When this computation is done for an ideal quantum gas (under the condition 
that the thermal wavelength is much smaller than the interparticle distance)
one finds that the quantum PF becomes the PF of the classical 
ideal gas. Evaluating the first quantum correction, one can appreciate 
that the quantum PF can be formally written as the PF a classical gas in which 
a fictitious two-body interaction term (the statistical potential) is 
added \cite{Pathria72,Huang87}. 
The statistical potential gives a simple characterization 
of the effects of the quantum statistics of the ideal gases: 
the statistical potential is ``attractive'' for bosons and ``repulsive'' 
for fermions, and it is respectively monotonically increasing (decreasing) 
for bosons (fermions). Another important result is that a suitable integral 
of the statistical interparticle potential gives the second coefficient 
of the virial expansion \cite{Pathria72,Huang87}. 

The statistical potential of a gas of ideal Abelian anyons has been studied 
in \cite{Khang92,Huang95,Huang07} and 
it depends on the statistical parameter $\alpha$ (we 
remind that $\alpha=0$ and $\alpha=1$ corresponds respectively 
to free two-dimensional spinless bosons and fermions, while 
$\alpha=1/2$ corresponds to semions \cite{Khare05}). 
It is found that for $1/2 \le \alpha \le 1$ the statistical potential 
$v_\alpha(r)$ is monotonically decreasing, while for $0<\alpha<1/2$ it has a 
a minimum at a finite value of $r$ and it is increasing for larger values 
of $r$ (while for $\alpha=0$ is monotonically increasing). We can refer 
to these behaviors respectively as ``quasi-fermionic'' 
($1/2 \le \alpha < 1$) and ``quasi-bosonic'' 
($0 < \alpha < 1/2$). The purpose 
of the present paper is to compute the statistical potential of the 
ideal gas of non-Abelian anyons: we find that the behaviour of the 
statistical potential depends on the Chern-Simons coupling and 
the isospin quantum number. As a function of these two parameters, 
quasi-bosonic and quasi-fermionic regions emerge, and they 
are part of a phase diagram which will be presented below.

The plan of the paper is the following: in Section II we recall 
the steps which lead to the computation of the statistics potential $v_\alpha$
of an ideal gas of Abelian anyons with statistical parameter $\alpha$. 
Since it is possible to show that the statistical potential for 
the non-Abelian gas may be written in terms of sums of Abelian 
statistical potentials (having different statistical parameters depending on the 
projection of the isospin quantum number), we provide in Section II a detailed 
study of $v_\alpha$: we present a compact and useful integral 
representation for it, showing that it can be written in terms of bivariate Lommel functions 
\cite{Watson}. Furthermore, using the Sumudu transform of the statistical potential 
\cite{Watugala93}, we also give a closed expression of $v_\alpha(r)$ in terms of the 
inverse Laplace transform of an algebraic function of $r$ and $\alpha$.  
Using this result we are able to give a simple expression for the 
statistical potential of the ideal gas of semions, and we 
show that for a general value $\alpha$ it is possible 
to retrieve the well-known result for the second virial coefficient 
of an ideal anyon gas found in \cite{Arovas85} 
(with an hard-core boundary condition 
for the two-body wavefunction at zero distance). For the sake 
of comparison with the non-Abelian case that follows, the limit behaviors both at small 
and large distance of the statistics potentials are presented. 
In Section III we introduce the non-Abelian Chern-Simons (NACS) 
model studied in the rest of the paper: we compute the statistics 
potential (within the hard-core boundary condition frame) as a function of 
the Chern-Simons coupling $\kappa$ and the isospin quantum number $l$ and 
we build a phase diagram summarizing the behavior of the statistical 
potential in terms of $\kappa$ and $l$. We then show that 
the second virial coefficient, previously studied in 
\cite{Lo93,Lee95,Hagen96,Mancarella12}, is 
correctly retrieved. The asymptotic expressions for the small 
and large distance of the statistics potentials are also given. 
Finally, our conclusions are drawn in Section IV, 
while supplementary material is presented 
in the Appendices.

\section{STATISTICAL POTENTIAL FOR ABELIAN ANYONS}
In this Section we introduce the model for an ideal gas of Abelian anyons, 
and we then derive its statistical potential $v_\alpha(r)$ as a function of 
the statistical parameter $\alpha$, obtaining 
the expression for $v_\alpha(r)$ given in 
\cite{Khang92,Huang95,Huang07}, and also providing an explicit formula 
for the semions (half-integer values of $\alpha$). 
The results for $v_\alpha(r)$ and the asymptotic expressions for small and large distance will be used in the next Section, where the statistical inter-particle potential of an ideal gas of non-Abelian anyons is derived and studied. 

Abelian anyons admit a concrete representation by the flux-charge composite model \cite{Khare05}, and the statistics of these objects can be understood in terms of Aharonov-Bohm type interference \cite{Aharonov59,Stern08}. The Hamiltonian for the quantum dynamics of 
an ideal system of anyons reads \cite{Lerda92,Khare05}
\begin{equation}
H_N=\sum_{n=1}^N \frac{1}{2M} \left(\vec{p}_n-\alpha\,\vec{a}_n\right)^2\,\,, 
\label{generalHamiltonian}
\end{equation}
where $\vec{p}_n$ is the momentum of the $n$-th particle 
($n=1,\cdots,N$). Similarly we will denote the position of 
the $n$-th particle by  $\vec{r}_n\equiv(x_n^1,x_n^2)$. 
In Eq.(\ref{generalHamiltonian}) $\alpha$ is the statistical parameter: notice 
that the physical quantities, e.g. the virial coefficients, are periodic 
with period $2$ \cite{Khare05}: 
the bosonic points are defined by $\alpha=2j$ and 
the fermionic ones by $\alpha=2j+1$, $j$ integer. For this reason we will consider 
in the following $\alpha\in[0,2]$. 

In Eq.(\ref{generalHamiltonian}) 
$\vec{a}_n$ is the vector potential carrying the flux attached to the bosons: indeed, the Hamiltonian (\ref{generalHamiltonian}) is written in the so-called bosonic representation, i.e. it is an Hamiltonian for identical bosons, and therefore acting on the subspace of wavefunctions which are symmetric with respect to the exchange of particles. The explicit expression for $\vec{a}_n\equiv(a_n^1,a_n^2)$ is 
\begin{equation}
a_n^i=\hbar \,\epsilon^{ij} \sum_{m (\neq n)} \frac{x^j_n-x^j_m}{\vert\vec{r}_n-\vec{r}_m\vert^2}\,\,. \label{fictitiouspotential}
\end{equation}
where $\epsilon^{ij}$ is the totally antisymmetric tensor ($i,j=1,2$). 

Let's recall that this model of Abelian anyons also admits a 
field-theoretic description: in fact (non-relativistic) 
anyons can be described by bosonic Schr\"odinger fields 
$\psi$ and $\psi^\dagger$ coupled to a Chern-Simons gauge field 
$a_\mu$ living in (2+1)-D 
\cite{HagenD31,phisrevd42} (then $\mu=0,1,2$). 
The Lagrangian density of such a system reads
$$
\mathcal{L}=\frac{c}{2} \, \epsilon^{\mu\nu\lambda}\,  a_\mu\partial_\nu a_\lambda + \psi^\dagger\left(i\,D_t+\frac{1}{2M}\vec{D}^2\right)\psi 
$$
where $c$ gives the measure of the interaction among particles mediated by the $U(1)$ gauge potential $a_\mu$, with 
the covariant derivatives given by $D_t=\partial_t+iq a_0$, $\vec{D}=\vec{\nabla}-iq\vec{a}$, and the anyon statistical parameter $\alpha$ to be identified 
as $\alpha \equiv q^2/(2\pi c).$ 

In the study of a quantum-mechanical ideal gas, the effect of the symmetry properties of the wave function can be interpreted, from a classical point of view, as a consequence of a fictitious classical potential introduced by Uhlenbeck and Gropper \cite{Uhlenbeck32}, referred to as effective statistical potential, which represents the first quantum correction for the classical PF \cite{Pathria72,Huang87}. 
For our purposes we have to consider the two-body case, which is relevant for the subsequent computation of the statistical interparticle potential \cite{Pathria72,Huang87}. The statistical potential completely determines the second virial coefficient, which gives the thermodynamical properties of the system in the dilute (high-temperature) regime. For the two-anyon system, after separating in (\ref{generalHamiltonian}) with $N=2$ 
the center-of-mass dynamics 
(i.e. that of a particle having mass $2M$), one is left with the dynamics of the relative wavefunction:
\begin{equation}
-\frac{1}{M}\left[\vec{\nabla}_r - i\,\alpha\, \vec{a}(\vec{r}) \right]^2\psi(\vec{r})=
E\,\psi(\vec{r})\,\,,
\label{COM}
\end{equation}
where $\vec{r}=\vec{r}_1-\vec{r}_2$ is the relative coordinate, and  $a^i(\vec{r})=\epsilon^{ij}\frac{x^j}{r^2}.$ Therefore the relative dynamics is equivalent
to that of a single particle in presence of a point vortex $\vec{a}(\vec{r})$
placed at the origin.

An analysis of the statistical interparticle potential for an ideal gas of Abelian anyons is presented in \cite{Khang92,Huang95}. 
The eigenfunctions of (\ref{COM}) reads
\begin{equation}
\Psi_\alpha= e^{i \mathbf{K}\cdot\textbf{R}}\,e^{i l\theta}\,J_{\vert l-\alpha \vert}(kr)\equiv e^{i \mathbf{K}\cdot\textbf{R}}\, \psi_\alpha\,\,, \label{eigenhardcore}
\end{equation}
where the capital (italic) letters respectively refer to center-of-mass (relative) coordinates. The bosonic description used in (\ref{generalHamiltonian}) imposes the condition $l=$ even; furthermore $J_m(x)$ denote the Bessel functions of the first kind \cite{Watson} (their definition is recalled in Appendix \ref{Bessel}). 
Notice that the wavefunctions (\ref{eigenhardcore}) are the eigenfunctions of (\ref{COM}) provided that the hard-core boundary conditions $\Psi_\alpha(0)=0$ are imposed.

\begin{figure}[t]
\vspace{0.4cm}
\centerline{\scalebox{0.8}{\includegraphics{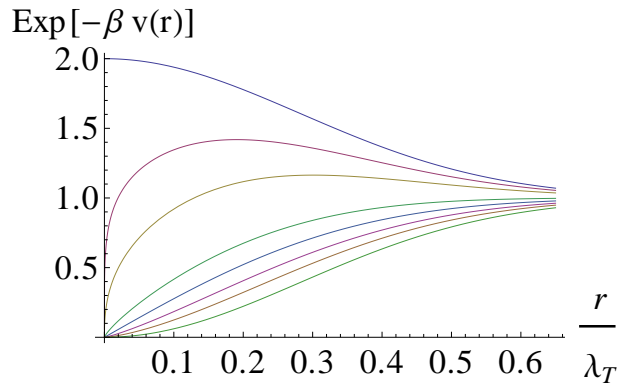}}}
\caption{Plot of $e^{-\beta v_\alpha(r)}$ vs. $r/\lambda_T$ for different values 
of the statistical parameter: from top to the bottom it is 
$\alpha=0,0.1,0.2,0.4,0.5,0.6,0.7,1$.}
\label{abelexpmbetav}
\end{figure}

The two-body PF is $Z=\Tr e^{-\beta H_2}=2A\lambda_T^{-2}Z',$ 
where $Z'$ is the single-particle PF in the relative
coordinates, $\beta=1/k_B T$, $\lambda_T=$ $ (\beta h^2 / 2 \pi M)^{1/2}$ is the thermal wavelength and $A$ is the area of the system. The relative PF $Z'$ is given by 
\begin{equation}
\label{Z_prima}
Z'=\frac{1}{h^2}\sum_{l=-\infty}^{\infty}\,\int d^2p \int d^2r\;e^{-\beta p^2/M}\,
\vert\psi_\alpha\vert^2
\end{equation}
(with $p=\hbar k$). 
It is possible to conveniently rewrite Eq.(\ref{Z_prima}) by using the following 
integral property \cite{Gradshteyn} of the Bessel functions:
\begin{equation}
\int_0^\infty e^{-\alpha x} J_\nu\left(2\beta\sqrt{x}\right)\,
J_\nu\left(2\gamma\sqrt{x}\right)dx=\frac{1}{\alpha}\,
I_\nu\left(\frac{2\beta\gamma}{\alpha}\right)\;e^{-(\beta^2+\gamma^2)/\alpha}\,\,,
\end{equation}
where $I_\nu$ is the modified Bessel function of the 
first kind \cite{Watson} (see also Appendix \ref{Bessel}). 
The relative PF then takes the form 
\begin{equation}
\label{partitionbessel}
Z'=\; \frac{1}{2}\sum_{n=-\infty}^{\infty}\,\int_0^\infty \,dx\,e^{-x}\,I_{\vert 2n-\alpha\vert}(x)\,\,,
\end{equation}
where 
\begin{equation}
\label{ICS}
x \equiv \frac{Mr^2}{2\beta \hbar^2}=\frac{\pi r^2}{\lambda_T^2}\,\,.
\end{equation} 
The one-body PF for classical systems, used in \cite{Uhlenbeck32} 
to define the concept of effective statistical potential, is
\begin{equation}
\label{abelpotpart}
Z'=\frac{1}{2 h^2}\int d^2p\, e^{-\beta p^2/M} \int d^2r\,e^{-\beta v(r)}\,\,:
\end{equation}
therefore, comparing (\ref{partitionbessel}) and (\ref{abelpotpart}) 
produces as a result \cite{Khang92}
\begin{equation} e^{-\beta v_\alpha(r)}=2\, e^{-x}\sum_{n=-\infty}^{\infty} I_{\vert 2 n-\alpha \vert}(x)
\label{potenzialestatisticokahng}
\end{equation} 
(see Appendix \ref{Statisticalpotential} for details).
Eq.(\ref{potenzialestatisticokahng}) is plotted in Fig.\ref{abelexpmbetav}.

For the considerations which follow, it is convenient to introduce the function
\begin{equation} 
\mathcal{M}_\alpha(x) \equiv \sum_{n=-\infty}^{\infty} I_{\vert 2 n-\alpha \vert}(x)\,\,, 
\label{def_M}
\end{equation}
so that the statistical potential can be written as
\begin{equation} 
e^{-\beta v_\alpha(r)}=2\, e^{-x} \, \mathcal{M}_\alpha(x)\,\,. 
\label{rel_M}
\end{equation}
The inter-particle statistical potential admits a closed expression 
in terms of the bivariate Lommel functions 
(alias, Lommel functions of two variables). Indeed, as evident from Appendix \ref{Bessel}, it is 
\begin{equation}\label{defineM}
\mathcal{M}_\alpha(x) = i^{-\alpha}\,U_{\alpha}(ix,ix)-i^{\alpha}\,U_{2-\alpha}(ix,ix)
\,\,,
\end{equation}
where $U_{\alpha}$ denote the Lommel functions 
of two variables \cite{Watson}. Notice that the symmetry property 
$v_\alpha(r)=v_{2-\alpha}(r)$ ($\forall \alpha \in \mathcal{R}$ ) holds for the statistical potential.

In Appendix \ref{Statisticalpotential} we prove the following integral representation for $v_\alpha(r)$, which will result useful in Subsection \ref{limitbeha} 
for discussing the large-distance limit behaviour:
\begin{equation}
\label{intreprbessel}
e^{-\beta v_\alpha(r)}=1+e^{-2 x} \cos{\alpha \pi}-2\,\frac{\sin{\alpha \pi}}{\pi} \,e^{-x} \,\int_0^\infty\,dt\,\frac{\sinh\left[(1-\alpha)t\right]}{\sinh t}\;e^{-x\,\cosh t}\,\,.
\end{equation}
Other integral representations for $e^{-\beta v_\alpha(r)}$ are presented 
in Appendix \ref{Statisticalpotential}. 
In the remaining parts of this Section we discuss the asymptotic 
behaviour of $v_\alpha$ for small and large values 
of the dimensionless parameter $x$, and we give a closed expression 
for the statistical potential $v_\alpha(r)$ 
in terms of the inverse Laplace transform 
of an algebraic function of $r$ and $\alpha$: this manipulation allows us 
to straightforwardly regain the known expressions for 
$v_\alpha(r)$ in the bosonic/fermionic cases, 
to obtain its expression in the case of semions and to 
finally recover the value of the second virial coefficient 
for a generic $\alpha$ presented in \cite{Arovas85}. 

\subsection{Limit behaviours}
\label{limitbeha}
In order to quantitatively understand the tendency of anyons to bunch together or vice versa to repel each other in different limit regimes of density, let's  
recall and further discuss the asymptotic behaviors of their effective statistical potential, for small and large distances \cite{Huang95,Huang07}.

For small $r$ (that is $x \ll 1$) we can approximate the summation term in 
(\ref{potenzialestatisticokahng}) as
\begin{equation}
\sum_{n=-\infty}^{\infty}I_{\vert 2 n-\alpha \vert}(x)\approx\sum_{n=0}^{\infty}
\left[\frac{1}{\Gamma(2n+\alpha+1)}(x/2)^{2n+\alpha}+\frac{1}{\Gamma(2n+3-\alpha)}(x/2)^{2n+2-\alpha}\right]
\approx \nonumber
\end{equation}
\begin{equation}
\label{head}
\approx[\Gamma(\alpha+1)]^{-1} (x/2)^{\alpha}+[\Gamma(3-\alpha)]^{-1} 
(x/2)^{2-\alpha}\,\,.
\end{equation}
Since
\begin{equation}
 \beta v_s(r)=-\ln\left[2 \;e^{-x}\sum_{n=-\infty}^{\infty}I_{\vert 2n-\alpha \vert}(x)\right]\approx
-\ln \left[2 \;e^{-x} \left([\Gamma(\alpha+1)]^{-1} (x/2)^{\alpha}+[\Gamma(3-\alpha)]^{-1} (x/2)^{2-\alpha}\right)\right]\,\,,
\end{equation}
it follows
\begin{equation} 
\beta v_\alpha(r) \approx \left\{ \begin{array}{lr}
-\ln\left[2 - 2 \pi r^2 /\lambda_T^2\right]\approx -\ln2+\frac{\pi}{2}(r/\lambda_T)^2,  & \alpha=0,\;2\\
-\ln\left[2 \; (\pi r^2 /2\,\lambda_T^2)^{\alpha}/\Gamma(\alpha+1)\right],  &  0<\alpha<1\\
-\ln\left[2 \pi r^2 /\lambda_T^2\right],  & \alpha=1\\
-\ln\left[2 \; (\pi r^2 /2\,\lambda_T^2)^{2-\alpha}/\Gamma(3-\alpha)\right], & 1<\alpha<2 \end{array}\right.
\end{equation}

We may summarize the small distance behaviour as follows: 
$v_\alpha(r)$ is repulsive and logarithmically divergent to 
$+\infty$ for any $\alpha \in (0,2)$, whereas for $\alpha=0,2$ it is attractive, 
and quadratically increasing in $r$ starting from the finite value $v_0(r=0)=-\ln 2$. 
The small-$r$ asymptotic function for $v_\alpha(r)$ is discontinuous in $\alpha=1$, 
being twice than the limit asymptotic functions for $\alpha \rightarrow 1^{\pm}$ (in fact two equal Bessel terms contribute to the asymptotic behaviour for $\alpha=1$, 
whereas only one of them dominates when $\alpha\neq 1$). 

To study the behavior of the statistical potential for large distance $r$ (i.e. $x\gg 1$) we employ the integral representation (\ref{intreprbessel}) of the modified Bessel function. The method of steepest descent allows to evaluate to an arbitrary order the last term of this integral representation. At the first significant order, we get 
\begin{equation}
-2\,\frac{\sin{\alpha \pi}}{\pi} \,e^{-z} \,\int_0^\infty\,dt\,\frac{\sinh\left[(1-\alpha)t\right]}{\sinh t}\;e^{-z\,\cosh t} \approx \frac{2(\alpha-1) \sin \alpha\pi}{\sqrt{2 \pi z}} e^{-2z}\,\,.\label{secondaddendterm}
\end{equation}
Therefore the large distance behavior of the statistical potential is given 
by
\begin{equation}
\beta v_s(r)\approx \left[-\cos \alpha\pi+
\frac{\sqrt{2}(1-\alpha)\sin\alpha\pi}{\pi r/ \lambda_T} \right] \,e^{-2 \pi r^2/ \lambda_T^2}
\end{equation}
for any $\alpha\in[0,2]$. Let us notice that this result differs from the corresponding one in \cite{Huang95}, and  that the asymptotic behaviours for $\alpha=0, \frac{1}{2}, 1$ (see formulas (\ref{laplacebosons}), (\ref{laplacefermions}), (\ref{laplacesemions}) in the sequel) are correctly retrieved. The statistical potential for large distance is vanishing for 
$r\rightarrow \infty$, and the interval $\alpha\in[0,1]$ 
(as well as the interval $\alpha\in[1,2]$, due to the 
symmetry property $v_\alpha=v_{2-\alpha}$) is divided in two regions: 
for large distance, $v_\alpha(r)$ is attractive for 
$0 \le \alpha < 1/2$, and repulsive for 
$1/2 \le \alpha \le 1$. 

\begin{figure}[t]
\vspace{0.4cm}
\centerline{\scalebox{0.8}{\includegraphics{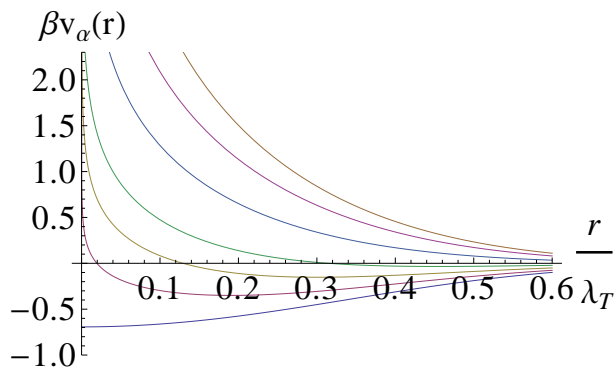}}}
\caption{$\beta v_\alpha(r)$ vs. $r/\lambda_T$ for different values of $\alpha$: 
from top to bottom it is $\alpha=1,0.7,0.5,0.3,0.2,0.1,0$. 
The bosonic curve ($\alpha=0$) is monotonically increasing while  
the fermionic curve ($\alpha=1$) is monotonically decreasing and 
divergent for $r \rightarrow 0$. 
All the curves for $0<\alpha<1/2$ diverge for $r \rightarrow 0$ and 
have a minimum point.}
\label{betavabeliani}
\end{figure}

\begin{figure}[t]
\vspace{0.4cm}
\centerline{\scalebox{0.8}{\includegraphics{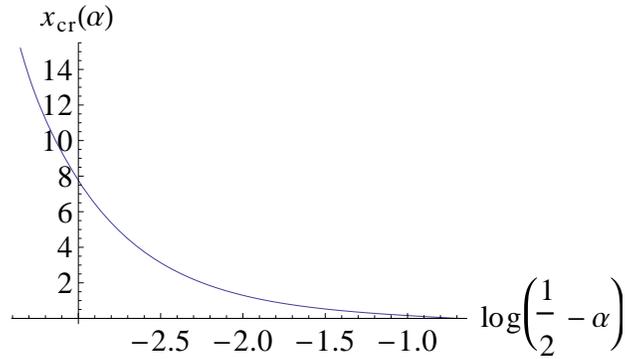}}}
\caption{The dimensionless value of the minimum point $x_{cr}(\alpha)$ 
of the statistical potential $v_\alpha$ of Abelian anyons 
as a function of the statistical parameter $\alpha$ near $\alpha=1/2$.}
\label{figminimum}
\end{figure}

The large-distance and short-distance behaviors, considered 
together, imply that $v_\alpha(r)$ must admit a minimum point 
at finite distance, $r_{cr}(\alpha)$ for any $0<\alpha<1/2$ 
(see Fig.\ref{betavabeliani}). We denote the corresponding dimensionless 
quantity by $x_{cr}(\alpha) \equiv \pi r_{cr}^2(\alpha)/\lambda_T^2$. 
The minimum point $x_{cr}(\alpha)$ tends to $+\infty$ for 
$\alpha\rightarrow \frac{1}{2}^-$, as shown in Fig.\ref{figminimum}. 

Fig.\ref{betavabeliani} clearly shows that Abelian anyons have, 
from the point of view of the statistical potential, a ``quasi-bosonic'' 
behavior for $0<\alpha<1/2$ (i.e., $v_\alpha(r)$ is non-monotonic with 
a minimum) and a ``quasi-fermionic'' behavior for $1/2 \leq \alpha<1$ 
($v_\alpha(r)$ is monotonically decreasing without a minimum).

\subsection{Laplace transform of the statistical potential and the $2^{nd}$-virial coefficient}
In this Section we write an explicit formula for $e^{-\beta v_\alpha(x)}$  
as the inverse Laplace transform of a function of $x$ and $\alpha$. 
This result will allow us to write a simple formula for the statistical 
potential for the semionic gas and to compute the second virial coefficient, 
which of course coincides with the result reported in the seminal reference 
\cite{Arovas85}.

Let's start by writing down the Sumudu transform \cite{Watugala93} of the 
function $\mathcal{M}_\alpha$, defined in Eq.(\ref{def_M}). The Sumudu 
transform of $\mathcal{M}_\alpha$ is defined as 
\begin{equation}
\label{SUMUDU}
\left[ \mathcal{M}_\alpha(x) \right]_S 
\equiv \int_0^{\infty} e^{-\theta}\mathcal{M}_\alpha(\theta x) d\theta=
\int_0^{\infty} e^{-t/x}\mathcal{M}_\alpha(t)\frac{dt}{x}=
s \int_0^{\infty} e^{-st}\mathcal{M}_\alpha(t) dt\,\,, 
\end{equation} where 
$s\equiv\frac{1}{x}$, whence 
\begin{equation}
\left[ \mathcal{M}_\alpha\left(\frac{1}{x}\right) \right]_S= x \, 
\mathcal{L}[\mathcal{M}_\alpha](x)\,\,,
\label{relazionesudumulaplace}
\end{equation} 
$\mathcal{L}$ being the ordinary (one-sided) Laplace transform. 
The function $\mathcal{M}_\alpha(x)$ is given by
\begin{equation}
\label{subst}
\mathcal{M}_\alpha(x)=\sum_{n=-\infty}^{\infty}I_{\vert 2 n-\alpha \vert}(x)=
e^{-\gamma \pi/2 i}\sum_{n=0}^{\infty}(-1)^n J_{2n+\gamma}(ix)+e^{-\alpha \pi/2 i}\sum_{n=0}^{\infty}(-1)^n J_{2n+\alpha}(ix)
\end{equation}
where $\gamma\equiv2-\alpha$. Substituting (\ref{subst}) in (\ref{SUMUDU}), 
one gets that the Sumudu transform of $\mathcal{M}_\alpha$ becomes 
$$\left[\mathcal{M}_\alpha(x)\right]_S=
e^{-\gamma \frac{\pi}{2} i} \sum_{n=0}^{\infty}(-1)^n \int_0^{\infty}e^{-t} J_{2n+\gamma}(ixt)dt
+e^{-\alpha \frac{\pi}{2} i} 
\sum_{n=0}^{\infty}(-1)^n \int_0^{\infty}e^{-t} J_{2n+\alpha}(ixt)dt\,\,.$$
The use of the integral properties of the Bessel functions of the first kind 
\cite{Watson} (see pg. 386) gives the following expression:
$$\left[ \mathcal{M}_\alpha(x) \right]_S=\frac{e^{-\gamma \frac{\pi}{2} i}}{\sqrt{1-x^2}}\sum_{n=0}^{\infty}(-1)^n \left(\frac{\sqrt{1-x^2}-1}{ix}\right)^{2n+\gamma}
+\frac{e^{-\alpha \frac{\pi}{2} i}}{\sqrt{1-x^2}}\sum_{n=0}^{\infty}
(-1)^n \left(\frac{\sqrt{1-x^2}-1}{ix}\right)^{2n+\alpha}= $$
$$=\frac{1}{\sqrt{1-x^2}}\left\{\left(\frac{(1-\sqrt{1-x^2})}{x}\right)^{\gamma}+\left(\frac{(1-\sqrt{1-x^2})}{x}\right)^{\alpha} \right\}
\frac{1}{1-\left(\frac{1-\sqrt{1-x^2}}{x}\right)^2}=$$ 
\begin{equation}
\label{intermediateexpression}
=\frac{1}{2}\left(\frac{1}{1-x^2}+\frac{1}{\sqrt{1-x^2}}\right)
\left[\left(\frac{(1-\sqrt{1-x^2})}{x}\right)^{\gamma}+
\left(\frac{(1-\sqrt{1-x^2})}{x}\right)^{\alpha} \right]\,\,.
\end{equation}
By sending $x\rightarrow 1/x$ in (\ref{intermediateexpression}), and applying (\ref{relazionesudumulaplace}), we obtain
$$ \mathcal{M}_\alpha\left(\frac{1}{x}\right)_S=\frac{1}{2}\left(\frac{x^2}{x^2-1}+\frac{x}{\sqrt{x^2-1}}\right) \left[\left(x-\sqrt{x^2-1}\right)^{\gamma}+
\left(x-\sqrt{x^2-1}\right)^{\alpha} \right] $$
and 
\begin{equation}\mathcal{L}[\mathcal{M}_\alpha](x)=\,\, \frac{\left(x-\sqrt{x^2-1}\right)^{1-\alpha}+
\left(x-\sqrt{x^2-1}\right)^{\alpha-1}}{2(x^2-1)}\,\,.
\end{equation}
Hence the inter-particle statistical potential admits the following form, for many purposes easier to handle than (\ref{potenzialestatisticokahng}), 
since it does not contain infinite sums: 
\begin{equation}
 e^{-\beta v_\alpha(r)}=e^{-x}
 \mathcal{L}^{-1}\left[\frac{\left(x-\sqrt{x^2-1}\right)^{1-\alpha}+\left(x-\sqrt{x^2-1}\right)^{\alpha-1}}{x^2-1}
\right]\,\,.\label{potenzialeconlaplace}
\end{equation}
The correct potentials for the bosonic and fermionic cases are 
straightforwardly reproduced: using known results for the Laplace transforms 
\cite{Bateman}, 
we get
\begin{equation} 
\label{laplacebosons}
e^{-\beta v_{\alpha=0}(r)}=e^{-x} \mathcal{L}^{-1}\left[\frac{2x}{x^2-1}\right]=e^{-x} \;2 \cosh x=1+e^{-2x}\,\,;
\end{equation}
\begin{equation}
\label{laplacefermions}
 e^{-\beta v_{\alpha=1}(r)}=e^{-x} \mathcal{L}^{-1}\left[\frac{2}{x^2-1}\right]=e^{-x}\; 2 \sinh x = 1-e^{-2x}\,\,.
\end{equation}
Eq.(\ref{potenzialeconlaplace}) gives a closed formula 
for the potential in the case of semions 
($\alpha=1/2$, or $\alpha=3/2$):  
\begin{equation}
\label{laplacesemions}
e^{-\beta v_{\text{sem.}}(r)}=e^{-x}
 \mathcal{L}^{-1}\left[\frac{\left(x-\sqrt{x^2-1}\right)^{1/2}+\left(x-\sqrt{x^2-1}\right)^{-1/2}}{x^2-1}\right]=
\mbox{erf}(\sqrt{2x})
\end{equation}
where $\mbox{erf}$ is the error function \cite{Clark}.

Eq.(\ref{potenzialeconlaplace}) allows us to recover the second virial 
coefficient of a gas made of identical Abelian $\alpha-$anyons, which is 
given by \cite{Arovas85}: 
\begin{equation} B_2(\alpha,T) =
\frac{1}{4}\lambda^2_T 
                  \left(-1 + 4\alpha - 2\alpha^2\right)\,\,.
\label{famousresult}
\end{equation}
The link between the $2^{nd}$-virial coefficient and the statistical potential  can be expressed in the form
\begin{equation}
B_2(\alpha,T)=\frac{\lambda_T^2}{2}\int_0^{\infty} dx\,[1-e^{-\beta v(x)}]
\,\,. \label{linkpotentialvirial}
\end{equation}

Using Eqs.(\ref{laplacebosons})-(\ref{laplacefermions})-(\ref{laplacesemions}), for the three special cases $\alpha=0$, $1$ and $1/2$ (corresponding 
respectively to bosons, fermions and semions) one immediately finds
\begin{equation}
B_2(\alpha=0,T)=\frac{\lambda_T^2}{2}\int_0^{\infty} dx\,[1-(1+e^{-2x})]=-\frac{1}{4} \lambda_T^2\,\,;
\end{equation}
\begin{equation}
B_2(\alpha=1,T)=\frac{\lambda_T^2}{2}\int_0^{\infty} dx\,[1-(1-e^{-2x})]=+\frac{1}{4} \lambda_T^2\,\,;
\end{equation}
\begin{equation}
B_2(\alpha=\frac{1}{2},T)=\frac{\lambda_T^2}{2}\int_0^{\infty} dx\,[1-\mbox{erf }(\sqrt{2x})]=
\frac{\lambda_T^2}{2}\int_0^{\infty} dy\,y\,[1-\mbox{erf }y]=\frac{1}{8} 
\lambda_T^2
\end{equation}
as it should be.

Finally,  the effective 2-body statistical potential written as in Eq. (\ref{potenzialeconlaplace})allows us to easily recover the expression of the
$2^{nd}$-virial coefficient even for a general statistical parameter $\alpha$. Indeed, by virtue of (\ref{potenzialeconlaplace}) and the dominated convergence theorem, one has
$$\frac{B_2(\alpha,T)}{\lambda_T^2} =\frac{1}{2}\int_0^{\infty} dx\,[1-e^{-\beta v(x)}]$$
$$=\frac{1}{2}\; \lim_{\varepsilon \rightarrow 0} \int_0^{\infty} dx\,\left[e^{-\varepsilon x}-e^{-(1+\varepsilon)x}\left(
 \mathcal{L}^{-1}\left[\frac{\left(x-\sqrt{x^2-1}\right)^{1-\alpha}}{x^2-1}
\right]-\mathcal{L}^{-1}\left[\frac{\left(x-\sqrt{x^2-1}\right)^{\alpha-1}}{x^2-1}
\right] \;\right) \;\right]$$
$$=\frac{1}{2}\; \lim_{\varepsilon \rightarrow 0}\left[\frac{1}{\varepsilon}-\mathcal{L}_{\vert
x=1+\varepsilon}\left(\mathcal{L}^{-1}\left[\frac{\left(x-\sqrt{x^2-1}\right)^{1-\alpha}}{x^2-1}
\right]-\mathcal{L}^{-1}\left[\frac{\left(x-\sqrt{x^2-1}\right)^{\alpha-1}}{x^2-1}
\right]\;\right)\;\right]$$

\begin{equation}\label{checkarovas}
=\frac{1}{2}\; \lim_{\varepsilon \rightarrow 0}\left[\frac{1}{\varepsilon}-\frac{(1+\varepsilon-\sqrt{2\varepsilon+\varepsilon^2})^{1-\alpha}+
(1+\varepsilon-\sqrt{2\varepsilon+\varepsilon^2})^{\alpha-1} }{2\varepsilon+\varepsilon^2}\right]
=\frac{1}{4}(-1 + 4\alpha - 2\alpha^2)\,\,, 
\end{equation}
that is just (\ref{famousresult}).

As a byproduct of Eqs. (\ref{rel_M}), (\ref{defineM}), (\ref{linkpotentialvirial}), 
(\ref{checkarovas}), we find an interesting integral property relevant to the bivariate 
Lommel functions (new, at the best of our knowledge):
\begin{equation}
\label{proprintlommel} 
\int_0^{\infty}\,dx\,\left\{1-2\,e^{-x}\left[\,i^{-\alpha}\,U_{\alpha}(ix,ix)-i^{\alpha}\,U_{2-\alpha}(ix,ix)\,\right]
\right\}=-\frac{1}{2}+2\alpha-\alpha^2\,\,.
\end{equation}

\section{STATISTICAL POTENTIAL FOR NON-ABELIAN ANYONS}
In this Section we discuss the statistical interparticle potential 
for a two-dimensional system of $SU(2)$ NACS spinless particles. 
The NACS particles are pointlike sources mutually interacting 
via a non-Abelian gauge field attached to them \cite{Dunne98}. 
As a consequence of their interaction, equivalent to a non-Abelian 
statistical interaction for a system of bosons, they are endowed with fractional spins and 
obey braid statistics as non-Abelian anyons.

Let's briefly introduce the NACS quantum mechanics 
\cite{Guadagnini90,Verlinde91,Lee93,Kim94,Bak94}. 
The Hamiltonian describing the dynamics of the $N$-body system of free 
NACS particles can be derived by a Lagrangian with a Chern-Simons term and 
a matter field coupled with the Chern-Simons gauge term \cite{Bak94}: the 
resulting Hamiltonian reads
\begin{equation} 
{H}_N=-\sum_{i=1}^{N} \frac{1}{M_i}\left(\nabla_{\bar 
z_i}\nabla_{z_i}  +\nabla_{z_i}\nabla_{\bar 
z_i}\right)
\label{hamiltonianadelmodello}
\end{equation}
where $M_i$ is the mass of the $i$-th particles, 
$\nabla_{\bar z_i}=\frac{\partial}{\partial \bar z_i}$ and 
$$
\nabla_{z_i}=\frac{\partial}{\partial z_i}  +\frac{1}{2\pi 
\kappa} \sum_{j\neq i} \hat Q^a_i \hat Q^a_j \frac{1}{ 
z_i -z_j}\,\,.
$$
In formula (\ref{hamiltonianadelmodello}) $i = 1, \dots, N$ 
labels the particles, $(x_i, y_i)=(z_i+\bar z_i, 
-i(z_i-\bar z_i))/2$ are their spatial coordinates, 
and $\hat Q^a$'s are the isovector operators which can be represented by some generators $T^a_l$ in a representation of isospin $l$ \cite{Lee93}.  The quantum number 
$l$ labels the irreducible representations of the group of the rotations induced by the 
coupling of the NACS particle matter field with the non-Abelian gauge field: as a consequence, the values of $l$ are of course quantized and 
vary over all the integer and the half-integer numbers, with $l=1/2$ being  
the smallest possible non-trivial value ($l=0$ corresponds to a system of free bosons). As usual, a basis of 
isospin eigenstates can be labeled by $l$ and the magnetic quantum number $m$ (varying in the range $-l,-l+1,\cdots,l-1,l$).

Hence the statistical potential depends in general on the value of the isospin quantum number $l$ and on the coupling $\kappa$ (and of course on the distance $r$ and the temperature $T$). 
The quantity $\kappa$ present in the covariant derivative
is a parameter of the theory. The condition $4 \pi \kappa \, = \, {\rm integer}$ 
has to be satisfied for consistency reasons \cite{Witten89, Guadagnini90}. In the following we denote for simplicity by $k$ the integer $4 \pi \kappa$.

For non-Abelian anyons, in analogy with (\ref{abelpotpart}), the effective statistical potential can be related to the relative PF according to the following expression: 
\begin{equation}
Z'_2(\kappa,l,T)-{Z'_2}^{(n.i.)}(l,T)=\frac{1}{2 h^2}\int d^2p\; e^{-\beta p^2/M} \int d^2r \left[ \exp[-\beta v(\kappa,l,r)]-
\exp[-\beta v^{(n.i.)}(l,r)]\right]\,\,,
\label{nonabelpartpot} 
\end{equation}
where $v^{(n.i.)}(l,r)$ refers to the system of particles with isospin $l$ and without statistical interaction ($\kappa \rightarrow \infty$). 
The potential $v^{(n.i.)}(l,r)$ can be 
expressed in terms of the potentials $v_{\alpha=0}(r)$ and $v_{\alpha=1}(r)$ for the free Bose and Fermi systems (endowed with the chosen 
wave-function boundary conditions). $Z'_2(\kappa,l,T)-{Z'_2}^{(n.i.)}(l,T)$ is the (convergent) variation of the divergent PF for the 
two-body relative Hamiltonian, between the interacting case in exam and the non-interacting limit $\kappa \rightarrow \infty$.   

For hard-core boundary conditions on the relative two-anyonic vectorial wave-function, the quantity $v^{(n.i.)}(l,r)$ which enters Eq.(\ref{nonabelpartpot}) 
is given by the projection onto the bosonic/fermionic basis:
\begin{equation}
e^{-\beta v^{(n.i.)}(l,r)}=\frac{1}{(2l+1)^2}\sum_{j=0}^{2l}\,(2j+1)\left[\frac{1+(-1)^{j+2l}}{2} \, e^{-\beta v_{\alpha=0}(r)} 
+\frac{1-(-1)^{j+2l}}{2} \, e^{-\beta v_{\alpha=1}(r)} \right]\,\,,
\end{equation}
in analogy with the procedure shown in \cite{Hagen96,Lee95,Mancarella12} for the computation of the $2^{nd}$-virial coefficient. By using the results 
$e^{-\beta v_{\alpha=0}(r)}=1+e^{-2x}$, $e^{-\beta v_{\alpha=1}(r)}=1-e^{-2x}$, 
one then obtains
\begin{equation}
\exp[-\beta v^{(n.i.)}(l,r)]=1+\frac{e^{-2x}}{2l+1}\,\,.
\end{equation}
Notice that this non-interacting quantity exactly corresponds to $(-1)^{2l}$ times the statistical potential for a system of identical $(2l)$-spin ordinary particles (fulfilling the spin-statistics constraint) at the same temperature, similarly to what argued in \cite{Mancarella12} about the $2^{nd}$-virial coefficient for the same system. 

In the interacting case (i.e. finite $k$), we can express 
the statistical potential in terms of statistical potentials of Abelian anyons:
\begin{equation}
e^{-\beta v(k,l,r)}=\frac{1}{(2l+1)^2}\sum_{j=0}^{2l}\,(2j+1)\left[\frac{1+(-1)^{j+2l}}{2} \, e^{-\beta v^B_{\omega_j}(r)} +\frac{1-(-1)^{j+2l}}{2} \, 
e^{-\beta v^F_{\omega_j}(r)} \right]\,\,,
\end{equation}
where $\omega_j\equiv [j(j+1)-2l(l+1)]/k$, and $v^B_{\omega_j}(r)$, $v^F_{\omega_j}(r)$ are the potentials for the Abelian $\omega_j$-anyon gases respectively in the bosonic and fermionic bases, given by 
\begin{equation}
\label{bbp}
e^{-\beta v^B_{\omega_j}(r)}=2 \, e^{-x} \mathcal{M}_{\omega_j}(x) 
\end{equation}
and 
\begin{equation}
\label{fbp}
e^{-\beta v^F_{\omega_j}(r)}=2\, e^{-x} \mathcal{M}_{\omega_j+1}(x)\,\,.   
\end{equation}
Both (\ref{bbp}) and (\ref{fbp}) are periodic quantities under the shift 
$\omega_j \rightarrow \omega_j+2$; it follows 
\begin{equation}
\label{nonabelpotlommel}
e^{-\beta v(k,l,r,T)}=\frac{2\,e^{-x}}{(2l+1)^2}\sum_{j=0}^{2l}\,(2j+1)\left[\frac{1+(-1)^{j+2l}}{2} \mathcal{M}_{\omega_j}(x) +\frac{1-(-1)^{j+2l}}{2} 
\mathcal{M}_{\omega_j+1}(x) \right] \,\,.
\end{equation}
Eq.(\ref{nonabelpotlommel}) gives the statistical potential for a gas of NACS particles.

\subsection{$2^{nd}$-virial coefficient}
An useful application (and check, at the same time) of Eq.(\ref{nonabelpotlommel}) consists in computing the second virial coefficient. The analogous of (\ref{linkpotentialvirial}) reads
\begin{equation}
B_2(k,l,T)=\frac{\lambda_T^2}{2}\int_0^{\infty} dx\,[1-e^{-\beta v(k,l,r)}]\,\,.
\end{equation} 
Substituting in its integrand both Eq.(\ref{nonabelpotlommel}) and the following decomposition of the unity
\begin{equation}
1=\frac{1}{(2l+1)^2}\,\sum_{j=0}^{2l}(2j+1)\left[\,\frac{1+(-1)^{j+2l}}{2}+\frac{1-(-1)^{j+2l}}{2}\,\right]\,\,,
\end{equation}
one obtains for $B_2(k,l,T)$
\begin{equation}
\frac{\lambda_T^2}{2(2l+1)^2}\sum_{j=0}^{2l}(2j+1) \int_0^{\infty} dx \left[\frac{1+(-1)^{j+2l}}{2}\,(1-2e^{-x}
\mathcal{M}_{\omega_j}(x))+\frac{1-(-1)^{j+2l}}{2}\,
(1-2e^{-x}\mathcal{M}_{\omega_j+1}(x)) \right]\,\,.
\end{equation}
By virtue of (\ref{proprintlommel}) one has then
$$
B_2(k,l,T)=\frac{\lambda_T^2}{(2l+1)^2}\sum_{j=0}^{2l}(2j+1)
\left[\frac{1+(-1)^{j+2l}}{2}\left(-\frac{1}{4}+
\gamma_j-\frac{1}{2}\gamma_j^2\right)+\frac{1-(-1)^{j+2l}}{2}
\left(-\frac{1}{4}+\eta_j-\frac{1}{2}\eta_j^2\right)\right]
$$
where $\gamma_j\equiv\,\omega_j\,mod\,2$ and 
$\eta_j\equiv\,(\omega_j+1)\,mod\,2$. This result matches with previous 
results reported in literature \cite{Hagen96,Mancarella12}, see in particular 
Eqs.(38) and (41) of \cite{Mancarella12}.

\subsection{Minimum points}
Using Eq.(\ref{nonabelpotlommel}), we can study if the gas of NACS has 
a ``bosonic-like'', ``quasi-bosonic'' or ``quasi-fermionic'' behaviour, according to its characterization in 
terms of the statistical potential. In correspondence with the analysis carried out 
in Subsection \ref{limitbeha}, we can address the problem of determining 
which points of the discrete parameter space $\{k,l\}$ 
are associated to the presence of an (interior) minimum point 
$r_{crit}(k,l,T)$ for the statistical potential $v(k,l,r,T)$ (which will be referred to as ``bosonic'' ones), which points correspond to a monotonically increasing $v(k,l,r,T)$ (``bosonic-like'' points) and which ones instead correspond to a $v(k,l,r,T)$ monotonically decreasing in $r$ (the ``quasi-fermionc'' ones). To this aim, let's exploit the limit behaviors of $\exp[-\beta v(k,l,r,T)]$ for small distance ($x \ll 1$) and large distance ($x \gg 1$), 
which straightforwardly arise from Eqs.(\ref{head}) and 
(\ref{secondaddendterm}). Notice that at $x=0$ one has 
\begin{equation}
e^{-\beta v(k,l,r=0)}=\frac{1}{(2l+1)^2}\sum_{j=0}^{2l}\,(2j+1)\left[(1+(-1)^{j+2l}) \, \delta(\gamma_j,0)+(1-(-1)^{j+2l}) \, \delta(\gamma_j,1)] \right]\,\,,
\end{equation}
where $\delta$ denotes the Kronecker delta function and $\gamma_j \equiv \omega_j mod\,2$. For large distance it is
\begin{equation}
e^{-\beta v(k,l,r\gg\lambda_T)}\approx \left\{ \begin{array}{lr}
1+\frac{e^{-2x}}{(2l+1)^2}\,s(j,k,l),  & \quad \text{if } s(j,k,l)\neq 0 \\
1-\frac{e^{-2 x}}{(2l+1)^2\sqrt{2\pi x}}t(j,k,l) ,  & \quad \text{otherwise}.
 \end{array}\right.\,\,,
\end{equation}
where 
$$s(j,k,l)\equiv \sum_{j=0}^{2l}\,(2j+1)(-1)^{j+2l}\cos{(\omega_j \pi)}$$ and 
$$t(j,k,l)\equiv \sum_{j=0}^{2l}\,(2j+1)\sin(\gamma_j \pi)\left[(1+(-1)^{j+2l})\,(\gamma_j-1)-(1-(-1)^{j+2l}) \,(\eta_j-1)] \right]\,\,$$ 
with $\gamma_j\equiv\,\omega_j\,mod\,2$, $\eta_j\equiv\,(\omega_j+1)\,mod\,2$.

\begin{figure}[t]
\vspace{0.4cm}
\centerline{\scalebox{0.4}{\includegraphics{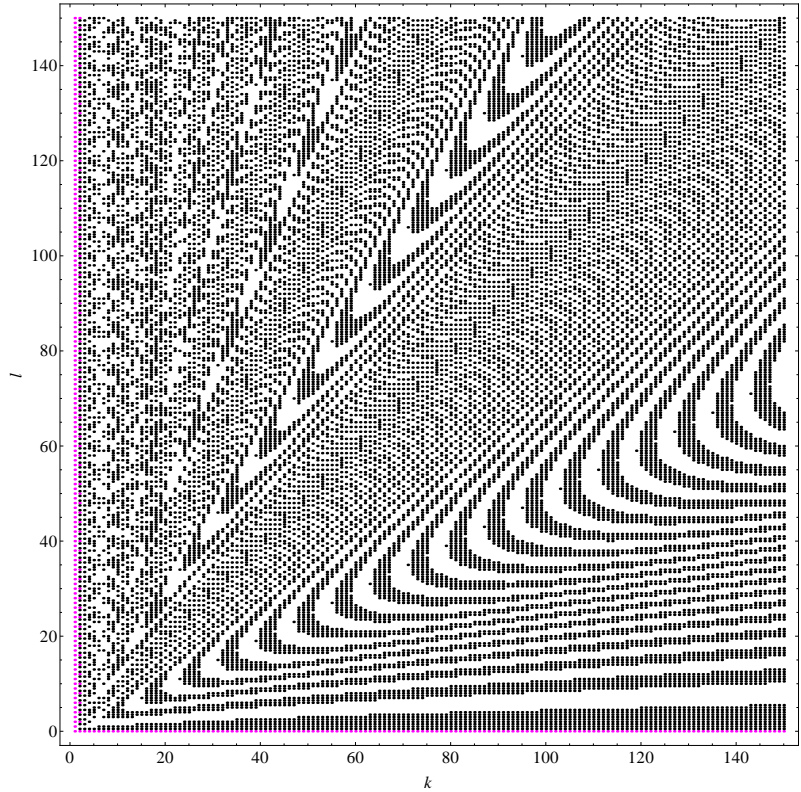}}}
\caption{Phase diagram in the parameter space $\{k,l\}$ 
for the family of non-Abelian anyon models under consideration. 
Points $(k,l)$ in the black regions correspond 
to the presence of an interior minimum point 
for the statistical potential $v(k,l,r,T)$ and are associated to 
quasi-bosonic behavior. Points in magenta correspond to a bosonic-like, monotonically increasing, behavior of the statistical potential. Finally, the remaining points are left blank and correspond to quasi-fermionic behavior.}
\label{phasediagram}
\end{figure}

Our results can be summarized in the ``phase diagram'' shown in  
Fig.\ref{phasediagram}, in which we distinguish pairs of parameters $(k,l)$ for which $v(k,l,r,T)$ has an interior minimum point (in black), pairs for which the statistical potential is monotonically increasing in $r$ (in magenta), and the remaining pairs (which are left blank). In this way the black points denote a ``quasi-bosonic'' behaviour and the magenta ones denote a ``bosonic-like'' behaviour, according to the classification operated in Section II to extract information from the statistical potential for Abelian anyons: one sees that for non-Abelian anyons one has mixed regions in which quasi-bosonic and quasi-fermionic 
behavior alternate, separated by regions dominated by a quasi-bosonic behavior. Bosonic-like behaviour occurs only for pairs $(k,l)$ having one of the forms: $(k \text{ generic},l=0)$, $(k=1,l\text{ integer})$, or $(k=2,l\text{ even})$.

\section{CONCLUSIONS}
In this paper we have studied the two-body effective statistical potential (which models, in the dilute regime, the dominant term of the statistical interaction between the particles) of ideal systems of Abelian and non-Abelian anyons, described within the picture of flux-charge composites. In both cases we have derived closed expression for the statistical potential 
(in terms of known special functions) and we have studied its behavior. Asymptotic 
expansions have been provided, and the second virial coefficients of both systems have been found using a compact expression of the statistical potential in terms of Laplace transforms. A phase diagram for the non-Abelian gas as a function of the Chern-Simons coupling 
and the isospin quantum number has been derived.

In our study we have considered hard-core boundary conditions for the relative 
anyonic wave-functions: however, it would be interesting to use the results obtained 
so far in order to analyze the statistical potential in the more general soft-core conditions 
(both for the Abelian and non-Abelian ideal anyon gases). 

\textit{Acknowledgements}: We wish to thank Andrea De Luca and 
Gianfausto Dell'Antonio for very valuable suggestions and discussions. 

\appendix

\section{Definition of the used Bessel functions}
\label{Bessel}
In the main text we used the Bessel functions of the first kind $J_\alpha$ and 
the modified Bessel function of the first kind $I_\alpha$: their definition 
is respectively given by
$$J_\alpha(x) = \sum_{m=0}^\infty \frac{(-1)^m}{m! \; \Gamma(m+\alpha+1)} \left(\frac{x}{2}\right)^{2m+\alpha}$$
and
\begin{equation}   I_\alpha(x) = i^{-\alpha} J_\alpha(ix) =\sum_{m=0}^\infty \frac{1}{m!\; \Gamma(m+\alpha+1)}\left(\frac{x}{2}\right)^{2m+\alpha} \,\,.\label{definemodified}
\end{equation}

The Lommel functions of two variables are defined in Eq.(5) of pg. 537 
of \cite{Watson} and read
$$U_\nu(w,z)=\sum_{m=0}^{\infty}(-1)^m\left(\frac{w}{z}\right)^{\nu+2m}J_{\nu+2m}(z)$$	
$$V_n(w,z)=\cos\left(\frac{w}{2}+\frac{z^2}{2w}+\frac{\nu \pi}{2}\right)+U_{2-\nu}(w,z)\,\,.$$

\section{Properties of the statistical potential $v_\alpha$}
\label{Statisticalpotential}
In this Appendix we provide details and further informations on the statistical 
potential $v_\alpha$ for Abelian anyons. 

We first give a derivation of Eq.(\ref{potenzialestatisticokahng}): 
the relative PF (\ref{partitionbessel}) is given by
\begin{equation}
Z'=
\frac{1}{2}\sum_{l=-\infty}^{\infty}\,\int_0^\infty \,dx\,e^{-x}\,I_{\vert l-\alpha\vert}(x),\quad x=M r^2 /2\beta\hbar^2
\end{equation}
and it can be rewritten as 
\begin{equation}
Z'=\frac{1}{2 h^2}\int d^2p \;e^{-\beta p^2/M} \int d^2r \sum_{l=-\infty}^{\infty}
2 e^{-M r^2/2\beta \hbar^2}\,I_{\vert l-\alpha\vert}\left(\frac{M r^2}{2\beta \hbar^2}
\right)\,\,,
\end{equation}
as pointed out in \cite{Khang92}. 
That allows for its comparison with the 
PF (\ref{abelpotpart}) 
for classical systems associated to a generic potential $v(r)$, 
whence the result (\ref{potenzialestatisticokahng}).

We also observe that 
\begin{equation}
\label{abellommel}
e^{-\beta v_\alpha(r)}=2\,e^{-x}\left[i^{-\alpha}\sum_{n=0}^{\infty}(-1)^n J_{2n+\alpha}(ix)+i^{\alpha-2}\sum_{n=0}^{\infty}(-1)^n J_{2n+2-\alpha}(ix)\right]\,\,.
\end{equation}
The expression that follows here below is a possible closed form for the statistical potential, but at the cost of using an integral representation 
given in formula (7), pg. 652 of \cite{Brychkov}. 
It stands for any complex number $\mu$ in the vertical strip $ -1 < \operatorname{Re} \mu <\alpha$: 
$$ e^{-\beta v_\alpha(r)}=2\,e^{-x}\left[i^{-\alpha}\sum_{n=0}^{\infty}(-1)^n J_{2n+\alpha}(ix)+i^{\alpha-2}\sum_{n=0}^{\infty}(-1)^n J_{2n+2-\alpha}(ix)\right]=$$
$$=2\,e^{-x}\left\{i^{-\alpha}\left[\frac{1}{2}\int_0^{ix} J_\mu(ix-t)\;J_{\alpha-\mu-1}(t)\;dt \right]+i^{\alpha-2}\left[\frac{1}{2}\int_0^{ix}
 J_\mu(ix-t)\;J_{1-\alpha-\mu}(t)\;dt \right]\right\}=$$
\begin{equation}
=e^{-x}\left[i^{-\alpha}\int_0^{ix} J_\mu(ix-t)\;J_{\alpha-\mu-1}(t)\;dt-i^{\alpha}\int_0^{ix} J_\mu(ix-t)\;J_{1-\alpha-\mu}(t)\;dt\right]\,\,.
\end{equation}
A simpler integral representation, valid for $\alpha\in(0,2)$, can be produced by using the following property 
(\cite{Watson}, pg. 540) of the bivariate Lommel function $U$:
$$U_\nu(w,z)=\frac{w^\nu}{z^{\nu-1}}\int_0^1\,J_{\nu-1}(zt)\,\cos\left\{\frac{1}{2}w(1-t^2)\right\}\,t^\nu\,dt,\quad \, \Re(\nu)>0$$
together with expressions (\ref{abellommel}), (\ref{defineM}) and 
(\ref{definemodified}). The resulting integral representation is  
\begin{equation}
e^{-\beta v_\alpha(r)}=2\,x\,e^{-x} \int_0^1\,\cosh\left[\frac{x}{2}(1-t^2)\right]\left(I_{\alpha-1}(xt)\,t^\alpha \,+\,I_{1-\alpha}(xt)\,t^{2-\alpha}\right)\,dt\,\,.
\end{equation}

In the final part of this Appendix we provide the derivation 
of the integral representation (\ref{intreprbessel}). To this end, 
we use the following representation  \cite{Gradshteyn} 
for the modified Bessel function of the first kind:
\begin{equation}
\label{reprintI}
I_\nu=\frac{1}{\pi}\int_0^\pi\,e^{z\,\cos\theta}\,\cos{\nu\theta}\,d\theta\,-\,\frac{\sin{\nu\pi}}{\pi}\int_0^\infty\,e^{-z\,\cosh t-\nu t}\,dt,\quad \arg \vert z\vert \leq \frac{\pi}{2},\, \Re \nu >0\,\,.  
\end{equation}
Then the summation term in (\ref{potenzialestatisticokahng}) is: 
\begin{equation}
\sum_{n=-\infty}^{\infty}I_{\vert 2 n-\alpha \vert}(z)=\sum_{n=0}^{\infty}I_{2n+2-\alpha}(z)+\sum_{n=0}^{\infty}I_{2n+\alpha}(z)=\nonumber
\end{equation}
\begin{equation}
=\frac{1}{\pi} \int_0^\pi d\phi\; e^{z \cos\phi} \left[\sum_{n=0}^{\infty} \cos{(2n+\alpha)\phi}+\sum_{n=1}^{\infty} \cos{(2n-\alpha)\phi}\right]-
\frac{1}{\pi} \int_0^\infty dt\;e^{-z \cosh t} f(t,\alpha)\,\,,
\label{secondotermine}
 \end{equation}
where
\begin{equation}
f(t,\alpha)=\sum_{n=0}^{\infty} e^{-(2n+\alpha)t} \sin{(2n+\alpha)\pi}+\sum_{n=1}^{\infty} e^{-(2n-\alpha)t} \sin{(2n-\alpha)\pi}=\sin\alpha\pi 
\frac{\sinh[(1-\alpha)t]}{\sinh t}\,\,,
\nonumber
\end{equation}
for $t\neq 0$ and 
\begin{equation}
f(0,\alpha) \equiv \lim_{t\rightarrow 0^{\pm}} f(t,\alpha)=(1-\alpha) \sin\alpha 
\pi\,\,.
\nonumber 
\end{equation}
The first addend of the last integral representation is 
{\small \begin{equation}
\frac{1}{\pi} \int_0^\pi d\phi\; e^{z \cos\phi} \left[\sum_{n=0}^{\infty} \cos{(2n+\alpha)\phi}+\sum_{n=1}^{\infty} \cos{(2n-\alpha)\phi}\right]=
\frac{1}{\pi} \int_0^\pi d\phi\; e^{z \cos\phi} \left[\sum_{n=0}^{\infty} \cos{(2n+\alpha)\phi} \,+ \right.
\nonumber
\end{equation}
\begin{equation}
+\left.\sum_{n=-\infty}^{-1} \cos{(2n+\alpha)\phi}\right]=
\frac{1}{\pi} \int_0^\pi d\phi\; e^{z \cos\phi}  \sum_{n=-\infty}^{+\infty}\cos{(2n+\alpha)\phi}=\frac{1}{\pi} \int_0^\pi d\phi\; e^{z \cos\phi} \; 
 \Re\left[ \sum_{n=-\infty}^{+\infty} (e^{i\alpha \phi} e^{2 i n\phi})\right]=
\nonumber
\end{equation}
\begin{equation}
\label{primotermine}
=\frac{1}{\pi} \int_0^\pi d\phi\; e^{z \cos\phi} \; \Re\left[e^{i\alpha \phi} \,2\pi\,\frac{\delta(2\phi)+\delta(2\phi-2\pi)}{2} \right]=\frac{1}{2}\,
(e^z+ e^{-z}\, \cos\alpha\pi)\,\,.
\end{equation}
As a result of (\ref{potenzialestatisticokahng}), (\ref{secondotermine}) 
and (\ref{primotermine}), one has then
\begin{equation}
e^{-\beta v_\alpha(r)}=1+e^{-2 z} \cos{\alpha \pi}-2\,\frac{\sin{\alpha \pi}}{\pi} \,e^{-z} \,\int_0^\infty\,dt\,\frac{\sinh\left[(1-\alpha)t\right]}{\sinh t}\;e^{-z\,\cosh t}\,\,.
\end{equation}
By direct inspection this result, 
notwithstanding the hypothesis of validity for (\ref{reprintI}), 
is valid also for $\alpha$ at the extremes of the interval $[0,2]$, 
so that the derivation of (\ref{intreprbessel}) is completed.

\end{document}